# Resonant inelastic x-ray scattering spectrometer with 25 meV resolution at Cu K-edge

Didem Ketenoglu,[1,2,†] Manuel Harder,[3,†] Konstantin Klementiev,[4] Mary Upton,[5]
Mehran Taherkhani,[1] Manfred Spiwek,[1]
Frank-Uwe Dill[1], Hans-Christian Wille[1] and Hasan Yavaş[1,*]

[1]*Deutsches Elektronen-Synchrotron DESY, Notkestraße 85, D-22607 Hamburg, Germany*
[2]*Ankara University, Faculty of Engineering, Department of Engineering Physics, Ankara, Turkey*
[3]*Technische Universität Dortmund, Lehrstuhl Experimentelle Physik I, Dortmund, Germany*
[4]*Max-IV Laboratory, Lund University, P.O Box 118, SE-22100 Lund, Sweden*
[5]*Argonne National Laboratory, 9700 S Cass Ave, Argonne, IL 60439 USA*
[*]hasan.yavas@desy.de

**Abstract:** An unparalleled resolution is reported with an inelastic x-ray scattering instrument at the Cu K-edge. Based on a segmented concave analyzer, featuring single crystal quartz ($SiO_2$) pixels, the spectrometer delivers a resolution near 25 meV (FWHM) at 8981 eV. Besides the quartz analyzer, the performance of the spectrometer relies on a four-bounce Si(553) high-resolution monochromator and focusing Kirkpatrick-Baez optics. The measured resolution agrees with the ray tracing simulation of an ideal spectrometer. We demonstrated the performance of the spectrometer by reproducing the phonon dispersion curve of a beryllium (Be) single crystal.

---

[†] These authors contributed equally.

## 1. Introduction

Resonant inelastic x-ray scattering (RIXS) has become a standard tool to probe elementary excitations in correlated electron systems. RIXS comes with unique advantages to study bosonic excitations across many Brillouin zones, including bulk sensitivity which results from the large penetration depth of high energy photons. Since the incoming photon energy is tuned to an absorption edge of the sample, the resonantly enhanced signal reveals element specific information [1–5].

RIXS is sensitive to a wide-range of excitations in correlated systems. It has been routinely used to study charge-transfer and d-d excitations in many transition metal compounds [6,7], particularly in the cuprate super conductors and their parent compounds [8,9]. However, due to the strong elastic peak in the K-edge RIXS spectra, the low-energy excitations like magnons and phonons had been hidden under the quasi-elastic tail. With the development of high-resolution RIXS instruments, detailed studies of bi-magnon excitations in $La_2CuO_4$ have become possible, albeit with longer data collection times due to still unfavorable signal to elastic background ratios [10–12]. Most recently, a glimpse of lattice excitations measured with Cu K-edge RIXS has also been reported [13].

High-resolution inelastic x-ray scattering (IXS) spectrometers utilize spherically bent crystal analyzers near back-scattering geometry on a Rowland circle [14]. Bending the analyzer crystal introduces further broadening in energy resolution due to the elastic deformation of the lattice planes. In order to mitigate this problem, the analyzer crystal is pixelated into a mosaic of unstrained small cubes [15–17]. With this arrangement, the energy resolution of the spectrometer is

$$\Delta E = \sqrt{(\Delta E_i)^2 + (\Delta E_c)^2 + (\Delta E_g)^2}, \qquad (1)$$

where $\Delta E_i$ is the resolution of the incident beam on the sample and $\Delta E_c$ is the intrinsic energy resolution of the analyzer crystal, which can be calculated using dynamical diffraction theory for perfect crystals [18]. $\Delta E_g$ is the geometrical contribution to the overall energy resolution, and can be denoted as

$$\Delta E_g = \frac{E}{R} \cot \theta_B \sqrt{(s)^2 + \left(\frac{d}{2}\right)^2}, \qquad (2)$$

where $E$ is the x-ray energy, $R$ is the diameter of the Rowland circle, $\theta_B$ is the Bragg angle on the analyzer crystal, and s is the effective beam size on the sample. $d$ is the pixel size of the position sensitive detector or the twice the cube size of the analyzer crystal if the detector does not have the position sensitivity [19,20].

The RIXS cross section depends significantly on the incident photon energy on the sample. The photon energy needs to be tuned across an absorption edge of the sample for the resonant effect; therefore, the spectrometer should be designed to perform around this resonant energy. Since the geometrical broadening $\Delta E_g$ dominates all the other factors as the Bragg angle recedes from 90° (Eq. 2), high-resolution RIXS measurements are constrained to the availability of back-scattering Bragg planes. Restricting analyzer fabrication to Ge and Si, as has effectively been done, has provided a very limited selection of available energies and resolutions. Non-cubic crystal structures with more than one atom in the unit cell, like sapphire [21,22] or quartz [23,24], provide an increase in the number of unique back-scattering planes to choose from[1]. The quartz (244) plane stands out as a good candidate for a high-resolution analyzer at the Cu K-edge energy with around 87° Bragg angle and close to 8 meV intrinsic energy resolution. This paper reports on simulations, development and testing of a new energy analyzer fabricated from quartz for measurements at the Cu K-edge.

---

[1] For a compilation of viable back-scattering reflections in Si, Ge, $LiNbO_3$, Sapphire, and Quartz: http://www.aps.anl.gov/Sectors/Sector27/AnalyzerAtlas/AnalyzerAtlas.html

## 2. Spectrometer

The spectrometer is stationed at the Dynamics Beamline - P01 of PETRA III at Deutsches Elektronen-Synchrotron, DESY. A liquid nitrogen-cooled double-crystal monochromator featuring Si(311) crystals is used to monochromatize the radiation from the undulator beamline, where Si(111) is also available as an option. A four-bounce Si (553) channel-cut monochromator in $(+--+)$ geometry is used for further monochromatization. A Kirkpatrick-Baez (KB) mirror system focuses the beam to a spot size of 3 µm x 9 µm (VxH) at the sample position, which is confirmed by knife-edge scans in both directions.

The analyzer, the sample, and the detector were placed on a Rowland circle with a 2 meter diameter, which can rotate around the sample to determine the momentum transfer to the sample. The geometry was arranged such that after being scattered from the sample in the horizontal direction, the x-ray beam was scattered from the analyzer in the vertical direction before being collected on the detector (Fig. 1).

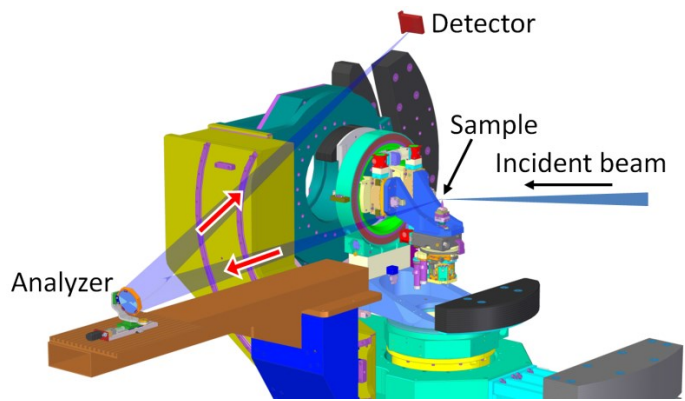

Fig. 1 The spectrometer geometry. The sample is at the center of rotation of a 6-circle diffractometer. The design of the diffractometer allows large space around the sample for the detector. The arrows show the propagation of the x-ray beam. A helium bag, which is not shown here, was put between the sample and the analyzer and the detector in order to avoid air absorption.

The quartz analyzer is placed 2 meters from the sample and the detector on a Rowland circle. The analyzer crystal has facets of 1.5 mm size and 0.2 mm gaps in between. The centers of the facets form a section of a sphere with 2 meter radius (see Appendix for the detailed fabrication procedures). The detector is a commercially available MYTHEN[2] 1D detector, which has 8 mm stripes with 50 µm pitch.

## 3. Ray tracing of the ideal spectrometer

The spectrometer was modelled by means of the ray tracing software xrt [25]. The goal of the modelling was to obtain the figures of energy resolution and energy acceptance, where all the contributing factors – spectrometer geometry, analyzer crystal shape and intrinsic resolution, beam size on the sample, and detector pixel size – are consistently taken into account.

The geometry of the photon source for the ray tracing mimicked the real experimental conditions. The beam size on the sample is taken as 3 µm x 9 µm (VxH) as it is mentioned above. The absorption of the primary and scattered beams in the sample was considered as negligible and therefore the intensity distribution over the source length was taken as uniform.

---
[2] https://www.dectris.com/mythen_overview.html

The response of the analyzer-detector pair to two distinct incident x-ray sources – a uniform energy distribution and a delta function – must be identified in order to determine the overall energy resolution of the spectrometer. For both cases, the width of the detector image $\Delta z$ is determined. The energy bandwidth of the analyzer-detector pair is

$$\delta\varepsilon = \frac{\Delta E_{uniform}}{\Delta z_{uniform}} \Delta z_{singleE} \, , \tag{3}$$

where $\Delta E$ is the energy bandwidth, and $\Delta z$ is the image size on the detector or the corresponding number of pixels. The size of the crystal facets of the analyzer along the dispersive direction determines the energy acceptance width $\Delta E_{uniform}$, which equals to 305.2 meV (FWHM) for the given facet size of 1.5 mm (see the zoomed footprint image in Fig. **2**). The energy acceptance width represents the energy range that is possible to analyze without any mechanical scanning, which is suitable for shot-by-shot measurements with emerging intense x-ray sources like free electron lasers.

The obtained ratio $\Delta E_{uniform}/\Delta z_{uniform}$ converts to the energy dispersion 5.087 meV/pixel, which was further used for the detector calibration. The image of a delta function source on the detector occupies ~3 pixels (3 x 5.087 ≈ 15.3 meV), which includes the intrinsic energy resolution of the quartz reflection (Fig. **2**).

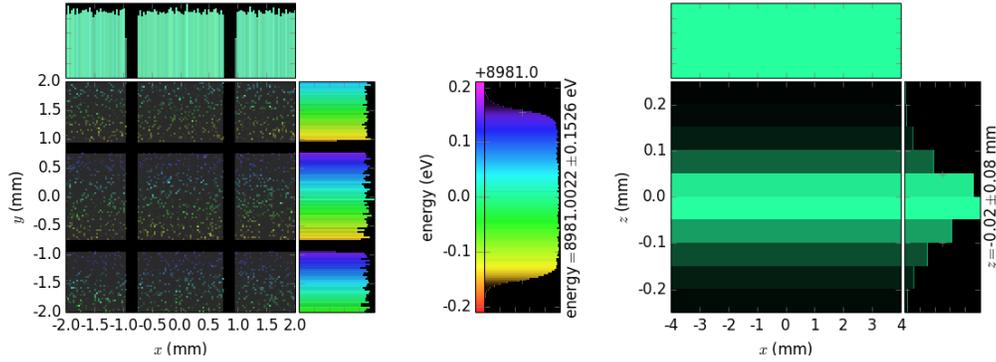

Fig. 2. Zoomed ray tracing image of the beam footprint over the quartz crystal facets (left panel). The pixel size of 1.5 mm by 1.5 mm and the groove size of 0.2 mm can be seen. The y-axis shows the dispersive direction, while the x-axis is non-dispersive. The middle panel shows the energy acceptance of the analyzer, i.e. response of the analyzer-detector pair to a uniform energy distribution. The right panel is the detector image of a delta function energy source.

The calculated energy resolution of the analyzer-detector system, $\delta E$ = 15.3 meV, was verified with ray tracing by considering a source energy distribution with seven lines (delta functions) spaced $\delta E$ apart. As shown in Fig. **3**, these seven lines are indeed spatially resolved. The definition of energy resolution adopted here gives a result similar to that given by the modified[3] Rayleigh criterion, which requires the local minima to be at most 80% of the maxima in order to be resolved [26].

---

[3] The *modified* criterion considers two lines as resolved if the intensity between them is reduced down to 80% relative to the maxima. This criterion is similar to the original Rayleigh criterion (the main maximum of one line is over the 1st minimum of the other line) and is used in cases which have no periodic minima.

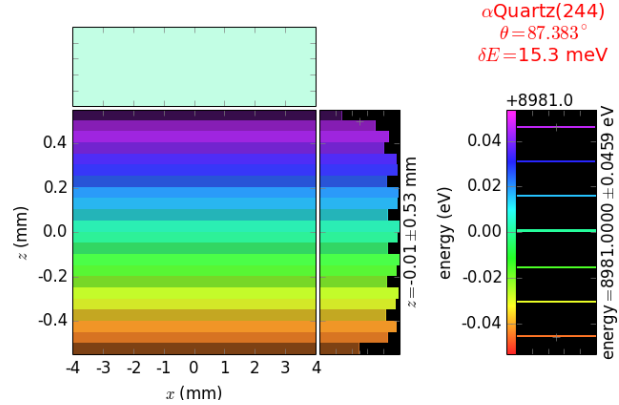

Fig. 3. The detector image of the 7-line source with the energy spacing equal to the calculated energy resolution (right panel). The 7-line source can be resolved by the spectrometer (left panel) since the local minima are around 80% of the maxima (modified Rayleigh criterion).

Finally, we modelled the response of the analyzer-detector pair to the energy bandwidth of the high-resolution monochromator (HRM). The ray tracing simulations of the Si (311) double crystal monochromator and the 4-bounce Si (553) HRM resulted in 16.4 meV, which was applied to the source seen by the analyzer. Fig. **4** shows the detector response to the 16.4-meV-wide energy bandwidth of the HRM from ray tracing. This image demonstrates the expected minimum width of the experimentally measured inelastic scattering spectra. It also justifies the selected detector pixel size (50 μm), as being sufficiently smaller than the narrowest measurable line width to not restrict total energy resolution. The overall energy resolution is calculated from the energy dispersion rate calculated earlier as 5 pixels x 5.087 meV/pixel ≈ 25.4 meV. A simple convolution of the bandwidth of the analyzer (15.3 meV) with the energy bandwidth of the HRM (16.4 meV) yields 22.4 meV, which does not include the discrete (pixelated) nature of the position sensitive detector.

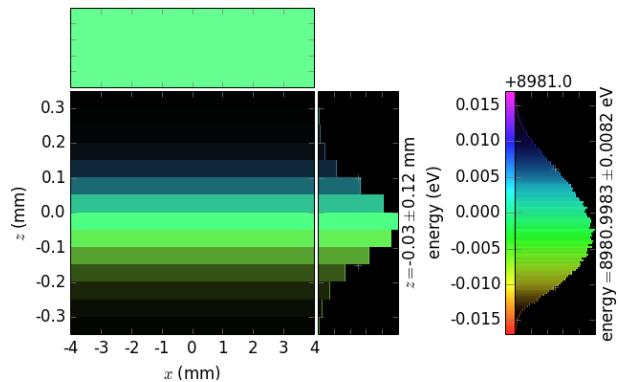

Fig. 4. The histogram of the counts on the detector from a ray tracing simulation. The discrete nature is due to the pixel size of the position sensitive detector (8 mm x 50 μm).

## 4. Performance of the spectrometer

In order to measure the highest possible resolution, we used two layers of Scotch® tape as a point-like elastic scatterer. The quartz analyzer of 100 mm diameter and 2 m radius of curvature was placed on a high-precision gimbal stage approximately 2 m from the sample (see Appendix for analyzer fabrication procedures and section 2 for the details of the spectrometer). The Rowland circle was optimized experimentally for the highest energy resolution by tweaking the analyzer-sample and analyzer-detector distance.

The analyzer's energy acceptance width of approximately 300 meV (Fig. **2**) was confirmed by scanning the incoming beam energy while keeping the analyzer-detector pair fixed. This step further confirmed the calibration of the analyzer-detector pair estimated by the ray tracing simulation as well as the dispersion rate of around 5 meV per pixel at 87° Bragg angle. The energy resolution is measured by scanning the incoming energy and recording a single detector pixel, which is cross-confirmed by passively collecting data without any mechanical scanning. The total energy resolution of the instrument was fitted with a Voigt function, which determined the full width at half maximum (FWHM) of the quasi-elastic line as $\Delta E_{tot}$ = 26meV (Fig. **5**). The measured value is only 3% larger than the ray tracing results. The reason for this discrepancy could be a number of things like a slight deviation from the Rowland geometry or residual stress due to the bonding between the analyzer pixels and the spherical backing. Considering the state-of-the-art resolution available at the MERIX instrument at the Advanced Photon Source (around 100 meV at the Cu K-edge energy) [19], this result is a major improvement for the Cu K-edge RIXS.

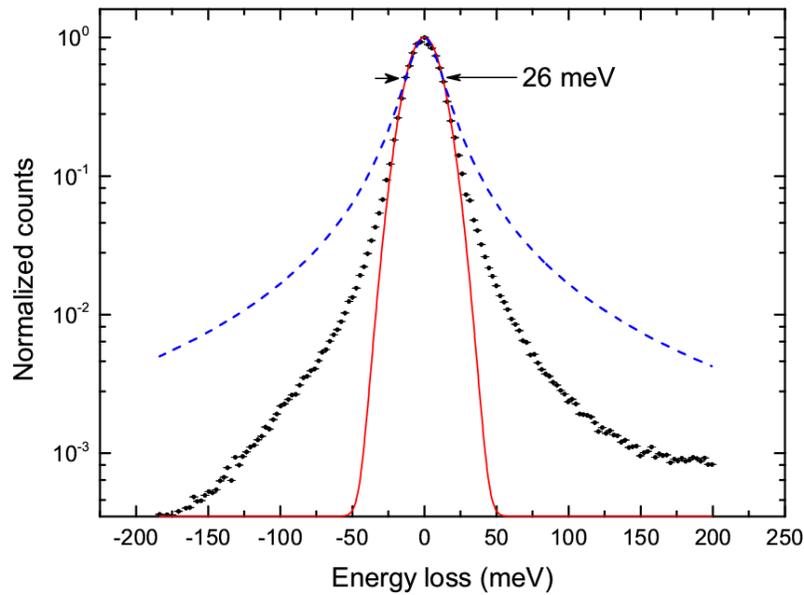

Fig. 5. Overall resolution function of the instrument. The data is collected by static counting on the position sensitive detector without any energy scanning. In order to have more data points on the peak, the Bragg angle is chosen to be 88.66° (2.624 meV/pixel). The FWHM (full width at half maximum) is measured to be 26 meV (with a Voigt function). A Gaussian (red solid line) and a Lorentzian (blue dashed line) with the same FWHM are shown for comparison. The y-scale is in logarithmic scale in order to show that the quasi-elastic tails are much better than a Lorentzian.

The performance of the analyzer was further tested by measuring the phonon dispersion of a beryllium single crystal due to its well-known dynamical structure. The spectra were collected in the first and the second Brillioun zones along (00ζ), and fitted with three peaks as was done with the previously developed sapphire analyzer [22] (Fig. 6). The detailed balance between the phonon creation and annihilation due to Boltzmann statistics was observed carefully. The resulting dispersion curve is in good agreement with previous measurements within experimental errors [27,28] (Fig. 7).

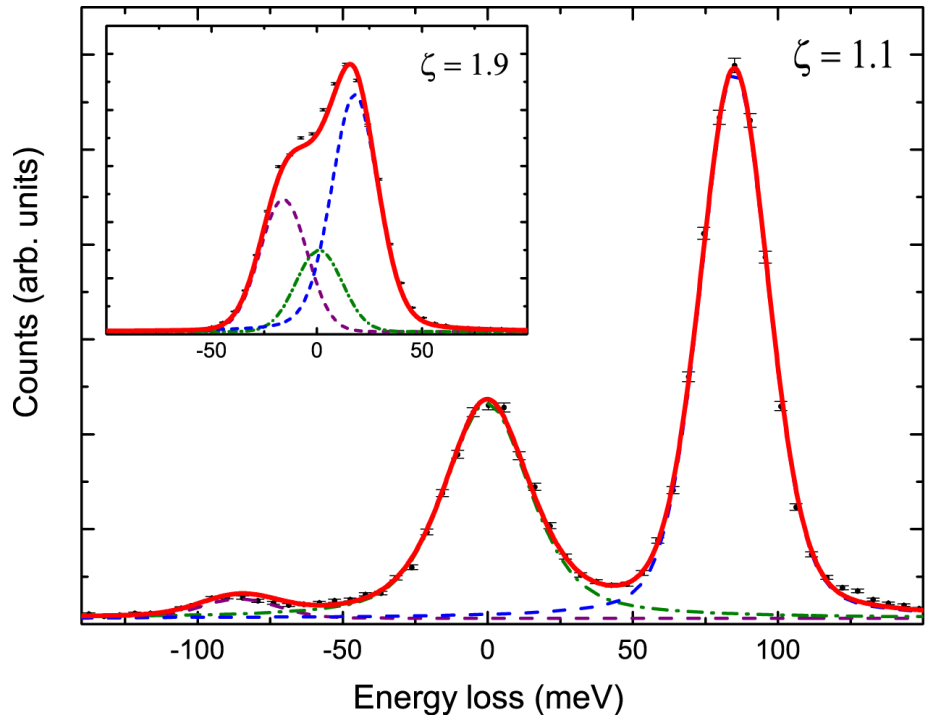

Fig. 6. Two representative spectra. The measurements were performed by statically collecting without any mechanical scans. The $\zeta$ denotes the momentum transfer values in units of $2\pi/a$, where a is the lattice parameter along (0 0 l). The peak in the middle (zero energy loss) is the elastic peak, and the peaks on the right and left are the phonon creation and annihilation peaks, respectively. Excitations at energies as low as 16 meV can be resolved (inner panel). Each spectrum was recorded in 300 seconds.

The current report confirms that the spectrometer's resolving power is good enough to resolve excitations at energies as low as 16 meV (Fig. 6). The instrument owes this record-breaking resolution to its quartz analyzer, which offers a suitable back-scattering Bragg plane for the Cu K-edge energy. It is expected that the many back-scattering planes available due to quartz's lower-symmetry structure, combined with the well-established processing methods described here will make quartz a viable material for high-resolution x-ray analyzers.

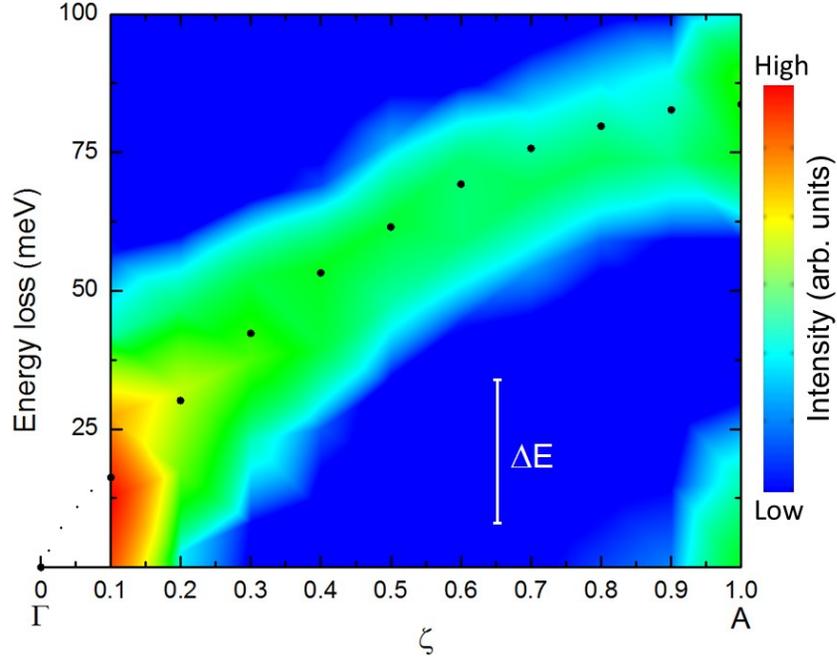

Fig. 7. Dispersion of phonons in Be single crystal along the [00ζ] in the first and second Brillouin zones. Black solid dots are from previous inelastic neutron scattering measurements [27,28]. The white line denotes the energy resolution of the current work.

**Appendix - Fabrication steps**

High-quality monolithic quartz wafers were grown, oriented and sliced by Tokyo Denpa Co. Ltd, and delivered by their German distributer Coftech GmbH. The wafer was diced with a high-precision diamond saw (DISCO Kiru-Kezuru-Migaku Technologies) with the depth of 500 μm leaving a back-wall of 1.5 mm. Un-diced surface of the crystal was coated with hydrofluoric acid resistant coating (HFRC) from Allresist GmbH to protect this side of the crystal before etching. The diced quartz was etched with 48% Hydrofluoric acid (HF) for 1 hour to remove the residual damage due to dicing. HFRC was cleaned off the un-diced surface after etching and the grooves were filled with HFRC. Piranha etch (Sulfuric acid-Hydrogen peroxide) was applied to make the surface of the 300 μm-thick silicon wafer hydrophilic before gluing to the diced surface of the quartz wafer using an epoxy (EPO-TEK® 301-2). The epoxy was cured at room temperature for 3 days. The quartz crystal was diced completely with the depth of 1.5 mm from the other side and HFRC was applied around the wafer, before 2 hours of final HF etching. HFRC was utilized to protect EPO-TEK® 301-2 which is not resistant to HF. The whole flat assembly was sandwiched between plano-concave and plano-convex glass substrates (Borofloat®-33) of 2 m radius. EPO-TEK® 301-2 was applied between the silicon wafer and the concave glass substrate. In order to avoid both air-bubble and dust in glue layer, gluing was performed in the clean room and extreme care was taken to distribute the glue uniformly. A thin layer of rubber was placed under the convex glass substrate to distribute the pressure uniformly and 75 μm-thick Kapton® foil was placed so that the rubber could be removed easily at the end. For bending process, approximately 1.8 tons were applied during the hardening of the epoxy. The procedure is summarized in Fig. 8.

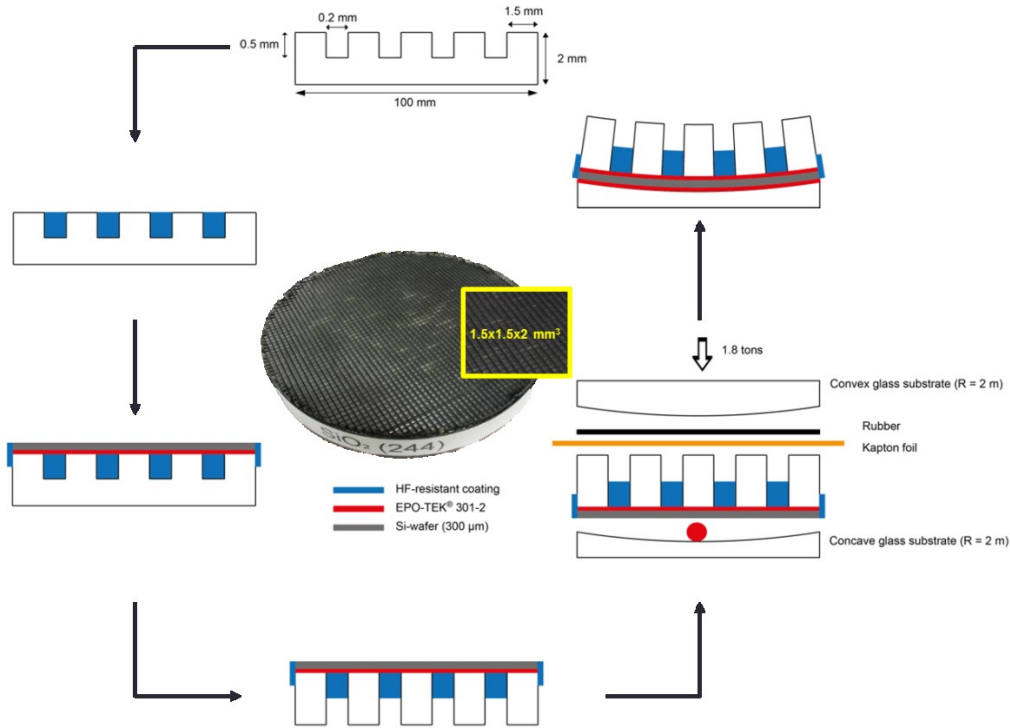

Fig. 8 The step-by-step procedure of fabricating a high-resolution quartz analyzer. Red dot and line depict the EPO-TEK glue. The blue color is the HF-resistant polymer. The picture in the middle is the finished analyzer.

**Acknowledgement**

Parts of this research were carried out at the light source PETRA III at DESY, a member of the Helmholtz Association (HGF). We would like to thank Milena Lippmann for her assistance in fabricating the quartz wafer; Andrey Siemens, Thorben Schmirander, and Jens Herda for their technical help at the beamline; Hermann Franz and the Photon Science management for continuing support; Horst Schulte-Schrepping, Ahmet Alatas, and Harald Sinn for stimulating discussions. DK acknowledges "The Scientific and Technological Research Council of Turkey (TUBITAK)" through "BIDEB-2219 Postdoctoral Research Fellowship". MH is supported by "Federal Ministry for Education and Research (Germany)" BMBF 05K13PE2k.